\documentclass[11pt]{article}
\usepackage{epsf} 
\usepackage{amsmath}
\usepackage{amssymb}
\usepackage{epsfig}
\usepackage{latexsym}
\usepackage{amsfonts}
\usepackage{graphicx}%
\usepackage{varioref}
\usepackage{ifthen}
\begin{document}
\parindent 0mm 
\setlength{\parskip}{\baselineskip} 
\thispagestyle{empty}
\pagenumbering{arabic} 
\setcounter{page}{0}
\mbox{ }
\rightline{UCT-TP-276/2010}
\newline
\rightline{MZ-TH/10-06}
\newline
\rightline{March  2010. REVISED May 2010}
\newline
%\vspace{0.1cm}

\begin{center}
{\huge \textbf{Chiral corrections to the ${\bf SU(2)\times SU(2)}$ Gell-Mann-Oakes-Renner relation}}
{\LARGE \footnote{{\LARGE {\footnotesize Supported in part by  the European FEDER and Spanish MICINN under grant FPA2008-02878, by NRF (South Africa), and by DFG (Germany)}}}}
\end{center}
\begin{large}
\begin{center}
{\bf J. Bordes}$^{(a)}$,
{\bf  C. A. Dominguez} $^{(b)-(c)}$,
{\bf  P. Moodley} $^{(b)}$, 
{\bf J.
Pe\~{n}arrocha}$^{(a)}$, 
and {\bf K. Schilcher} $^{(d)}$ \\
\end{center}
\end{large}
\begin{center}
$^{(a)}$Departamento de F\'{\i}sica Te\'{o}rica,
Universitat de Valencia, and Instituto de F\'{\i}sica Corpuscular, Centro
Mixto Universitat de Valencia-CSIC\\
\vspace{.4cm}
$^{(b)}$Centre for Theoretical Physics \& Astrophysics, University
of Cape Town, Rondebosch 7700, South Africa\\
\vspace{.4cm}
$^{(c)}$Department of Physics, Stellenbosch University, Stellenbosch 7600, South Africa\\
\vspace{.4cm}
$^{(d)}$Institut f\"{u}r
Physik, Johannes Gutenberg-Universit\"{a}t, Staudingerweg 7, D-55099
Mainz, Germany
\end{center}

\begin{center}
\textbf{Abstract}
\end{center}
\noindent
The next to leading order chiral corrections to the $SU(2)\times SU(2)$ Gell-Mann-Oakes-Renner (GMOR) relation are obtained using the pseudoscalar correlator to five-loop order in perturbative QCD, together with  new finite energy sum rules (FESR) incorporating polynomial, Legendre type, integration kernels. The purpose of these kernels is to suppress hadronic contributions in the region where they are least known. This reduces considerably the systematic uncertainties arising from the lack of direct experimental information on  the hadronic resonance spectral function. Three different methods are used to compute the FESR contour integral in the complex energy (squared) s-plane, i.e. Fixed Order Perturbation Theory, Contour Improved Perturbation Theory, and a fixed renormalization scale scheme. We obtain for the corrections to the GMOR relation, $\delta_\pi$, the value $\delta_\pi = (6.2\, \pm \,1.6) \%$.  This result is substantially  more accurate than previous determinations based on QCD sum rules; it is also more reliable as it is basically free of systematic uncertainties. It implies a light quark condensate $\langle0|\bar{u} u|0\rangle \simeq \langle0|\bar{d} d|0\rangle \equiv \langle0|\bar{q} q|0\rangle|_{2\,\mathrm{GeV}} = (- 267 \pm 5\; \mbox{MeV})^3$.
As a byproduct, the chiral perturbation theory (unphysical) low energy constant $H^r_2$ is predicted to be $H^r_2 (\nu_\chi = M_\rho) = - (5.1 \,\pm\, 1.8)\,\times 10^{-3}$, or  $H^r_2 (\nu_\chi = M_\eta) = - (5.7 \,\pm\, 2.0)\,\times 10^{-3}$.\\

%\bigskip KEYWORDS: QCD, Sum Rules.
\newpage

\section{Introduction}

The method of QCD sum rules\cite{REVIEW} provides a successful analytical technique to determine hadronic parameters and form factors, as well as to extract from data fundamental parameters of QCD, such as quark masses and the strong coupling. The method is based on two fundamental pillars, (i) the Operator Product Expansion (OPE) of current correlators at short distances, extended beyond perturbation theory, and (ii) Cauchy's theorem in the complex energy (squared) plane, also known as quark-hadron duality. The extension of the OPE effectively accounts for quark-gluon confinement by parameterizing propagator corrections in terms of a set of vacuum matrix elements of the quark and gluon fields entering the QCD Lagrangian, i.e. the so called vacuum condensates.\\
In this paper we concentrate on the pseudoscalar current correlator with pionic quantum numbers, i.e. the two-point function involving the up- and down-quark degrees of freedom. Specifically, we will study this correlator at zero momentum, which through a Ward identity leads to the $SU(2)\times SU(2)$ Gell-Mann-Oakes-Renner relation \cite{GMOR}-\cite{BROAD2}
%Eq.1
\begin{equation}
\psi _{5}(0)\,\equiv \,-(m_{u}+m_{d})\left\langle 0 \right\vert 
\overline{u}u+\overline{d}d\left\vert 0 \right\rangle\, = \,2\,f_\pi^2\,M_\pi^2\,(1\,-\,\delta_\pi)\;,
\end{equation} 
where $f_\pi = 92.21 \pm 0.14\;\mbox{MeV}$ \cite{PDG}-\cite{ROSNER}.
Our goal is to determine the corrections to this relation, i.e. $\delta_\pi$. In addition to its intrinsic importance to QCD, this quantity is of great interest to chiral perturbation theory (CHPT), as it is related to two of its low energy constants, as well as to the corrections to the $SU(3)\times SU(3)$ GMOR relation, i.e. at next to leading order \cite{BERN1}-\cite{BERN2}
%Eq.2
\begin{equation}
\delta_\pi = 4 \;\frac{M_\pi^2}{f_\pi^2} ( 2 L^r_8 - H^r_2) \;\;\;\; \mbox{and} \;\;\;\; \delta_K = \frac{M_K^2}{M_\pi^2} \,\delta_\pi \;,
\end{equation}
where $L^r_8$ is a physical low-energy constant, and $H^r_2$ an unphysical one.
The use of QCD sum rules to determine the chiral corrections to the GMOR relation has been fully justified in QCD in \cite{BROAD2}, \cite{JAMIN1}-\cite{CAD1} (for a recent reassessment see \cite{DNS1}). 
If one were to have direct experimental data on the hadronic spectral function (beyond the pion pole), the task of determining $\delta_\pi$ would be rather straightforward. Unfortunately, one only knows from experiment the mass and width of the first two radial excitations of the pion. This information is hardly enough to reconstruct the full hadronic spectral function. In fact, inelasticity, non-resonant background, and resonance interference are difficult, if not impossible to guess. Many attempts have been made in the past to model this spectral function, starting with threshold constraints from CHPT imposed on Breit-Wigner forms \cite{CAD2}. However, a serious systematic uncertainty is unavoidable. Historically, this has affected the accuracy of results from analyses based on the pseudoscalar correlator, such as e.g. QCD sum rule determinations of (light) quark masses \cite{REVIEW}, and estimates  of the corrections to the GMOR relation \cite{JAMIN2}, \cite{CAD3}. Recently, a major breakthrough has been achieved by introducing analytic integration kernels in QCD Finite Energy Sum Rules (FESR) tuned to suppress the (unknown) contribution of the resonance sector \cite{DNS1}, \cite{DNS2}-\cite{DNS3}. In this way it has been possible to reduce this contribution to a few percent of the total, which is made up by the pion (kaon) pole plus perturbative and non-perturbative QCD. Results for the up-, down-, and strange-quark masses, as well as the ratio of the strange to non-strange quark condensate are now available with basically no systematic uncertainty from the hadronic resonance sector.\\

In \cite{DNS1} the correction $\delta_\pi$ was obtained indirectly by determining the ratio of the strange to the non-strange quark condensates, $R_{su} \equiv \langle 0|\bar{s}s|0 \rangle/\langle 0|\bar{u}u|0 \rangle$,  together with the CHPT estimate \cite{JAMIN2}  $L^r_8 (\nu_\chi = M_\rho)= (0.88 \,\pm\, 0.24)\, \times\,10^{-3}$. The integration kernel used in the QCD FESR was chosen as a second degree polynomial constrained to vanish at the peaks of the two radial excitations of the ground state pseudoscalar meson. The method of Fixed Order Perturbation Theory (FOPT) was used to compute the FESR contour integral in the complex energy (squared) plane. 
The results thus obtained are \cite{DNS1}
%Eq.3
\begin{equation}
H^r_2(\nu_\chi = M_\rho) = - (4.3\,\pm\, 1.3)\,\times 10^{-3}\;\;\;\;\;\;\;\mbox{and} \;\;\;\;\;\;\;\;\delta_\pi = (4\,\pm\,2)\,\%.
\end{equation}
In this work we determine $\delta_\pi$ directly from a QCD FESR involving Legendre-type polynomials as integration kernels. These kernels are tuned to reduce considerably the importance of the hadronic resonance contribution, thus reducing the systematic uncertainty arising from  the lack of direct data in this region. This tuning is done in two ways, (i) as in \cite{DNS1}, \cite{DNS2}-\cite{DNS3}, i.e. using a second degree polynomial constrained to vanish at the peaks of the two radial excitations of the pion (local constraint), and (ii) using more general Legendre type polynomials
with global, rather than local, constraints. 
This direct determination  leads to a considerably more accurate result for $\delta_\pi$, i.e. an estimated total uncertainty of 25 \%, rather than 50 \% as in \cite{DNS1}, with both results being in agreement within errors.

\section{Finite energy sum rules and the Gell-Mann-Oakes-Renner relation}

We begin by introducing the light quark pseudoscalar correlator
%Eq.4
\begin{equation}
\psi _{5}(s=-q^{2})\,=\,i\int \,d^4 x\,e^{iqx}<0
|\,T(j_{5}(x)\,j_{5}(0))\,|0 >\;,
\end{equation}
where $<0|$ is the physical vacuum and the current density $j_{5}(x)$ is
%Eq.5
\begin{equation}
j_{5}(x)\,=\,(m_{d}+m_{u})\;\overline{d}(x)\,i\,\gamma _{5}\,u(x)\ ,
\end{equation}
and $m_{u,d}$ are the light quark masses. Invoking Cauchy's theorem in the complex energy (squared) s-plane, the correlator at zero momentum is given by
%Eq.6
\begin{equation}
\psi_5(0) \Delta_5(0) =  \frac{1}{\pi }\int_{s_{th}}^{s_0} \frac{\Delta_5(s)}{s}\,Im\,\psi _{5}(s)\; ds
\, +
\frac{1}{2\pi i}\oint_{C(|s_0|) } \,\frac{\Delta_5(s)}{s} \;\psi _{5}(s)
\,ds\,,
\end{equation}
where $s_{th}\, =\,  M_\pi^2$ is the hadronic threshold, the integration path $C(|s_0|)$ is a circle of (finite) radius $s_0$, and $\Delta_5(s)$ is an arbitrary, but analytic integration kernel. The radius $s_0$ determines the onset of the continuum, which is usually assumed to be given by perturbative QCD (PQCD). Hence, the first integral in Eq.(6) involves the hadronic spectral function, and the second integral the QCD correlator.
Saturation of the hadronic integral with only the pion pole  leads to the GMOR relation, Eq.(1) with $\delta_\pi=0$, after invoking the Ward identity
%Eq.7
\begin{equation}
\psi _{5}(0)\,=\,-[m_{u}(\mu^2)+m_{d}(\mu^2)]\left\langle 0 \right\vert 
\overline{u}u+\overline{d}d\left\vert 0 \right\rangle (\mu^2)\,.
\end{equation}
If quartic quark mass corrections to $\psi _{5}(s)$ are neglected, quite legitimate for the up and down quarks, the condensate in Eq.(7) can be considered normal-ordered (for a discussion about these corrections see \cite{JAMIN1}, \cite{CAD1}). Both the quark masses and the normal ordered vacuum condensates depend on the renormalization point $\mu^2$,
but their product does not.\\

In the QCD sector the pseudoscalar correlator has been known for quite some time   up to fourth loop order in PQCD \cite{CHET0}, and more recently the fifth order has been calculated \cite{CHET}. Quartic quark mass corrections, as well as vacuum condensate terms are also known, but we found their contribution to be negligible in this application. In fact, this contribution is much smaller than the uncertainty due to the value of the strong coupling. Hence, the two-point function entering the contour integral in Eq.(6) can be generically written as 
%Eq.8
\begin{equation}
\psi _{5}(s)|_{PQCD}= - \left[ m_{u}(\mu^2)+m_{d}(\mu^2)\right]^{2}\,%
\frac{3}{8\pi ^{2}}\; s\,\sum_{i=0}^{4}\psi _{5}^{(i)}(s,\mu^2)\,\,\left( \frac{
\alpha _{s}(\mu^2)}{\pi }\right) ^{i},  
\end{equation}%
where
%Eq.9
\begin{equation}
\psi _{5}^{(i)}(s,\mu^2)=\sum_{j=0}^{i+1}c_{ij}\,\,\left( \ln \frac{-s}{\mu ^{2}}%
\right) ^{j}\ \ \ \ \ (i=0,\ldots 4)  
\end{equation}%
and the constant coefficients $c_{ij}$ are given in \cite{CHET0}-\cite{CHET}.

The hadronic spectral function can be split into the pion pole term followed by the resonance contribution, i.e.
%Eq.10
\begin{equation}
\frac{1}{\pi} \, Im \;\psi _{5}(s)\,=\, 2\, f_\pi^2 \,M_\pi^4\, \delta(s - M_\pi^2)\; + \; \frac{1}{\pi} \, Im \;\psi _{5}(s)|_{RES} \;,
\end{equation}
where the resonance contribution, $Im \;\psi _{5}(s)|_{RES}$, involves the radial excitations of the pion, with the first two, $\pi(1300)$ and $\pi(1800)$, known from experiment \cite{PDG}. This experimental information, though, is restricted to the values of the mass and the width. The resonance spectral function in this channel is completely unknown experimentally, so that its reconstruction is model dependent. This leads to a serious systematic uncertainty. In an attempt to reduce  this model dependency, it was first proposed long ago \cite{CAD2} to normalize this spectral function at threshold using  CHPT as a constraint.
%Eq.11
\begin{equation}
\frac{1}{\pi}\; Im \;\psi_5(s)|_{\pi\pi\pi} \; = \theta(s \,- \,9\,M_\pi^2) \;\frac{1}{9}\;\frac{M_\pi^4}{f_\pi^2} \frac{1}{2^8\,\pi^4}\; I|_{PS}(s)\;,
\end{equation}
where the phase-space integral $I|_{PS}(s)$ is given by \cite{RD}
%Eq.12
\begin{eqnarray}
I|_{PS}(s)\;&=&\; \int_{4 M_\pi^2}^{(\sqrt{s} - M_\pi)^2}\; du \;\sqrt{1 - \frac{4M_\pi^2}{u}} \; \;\lambda^{1/2}(1,u/s,M_\pi^2/s)\;
\Biggl\{5 + \frac{1}{2} \;\frac{1}{(s-M_\pi^2)^2}
\Biggr. \nonumber \\ [.3cm]
&\times& \left.   \Bigl[(s - 3 u + 3 M_\pi^2)^2 
+  3\; \lambda(s,u,M_\pi^2) \;\left(1 - \frac{4\, M_\pi^2}{u}\right) + 20 \;M_\pi^4 \Bigr]  \Biggr. \right.\nonumber \\ [.3cm]
&+& \Biggl. \frac{1}{(s - M_\pi^2)}\; \Bigl[ 3 (u - M_\pi^2) - s + 9 M_\pi^2 \Bigr] \Biggr\} \;,
\end{eqnarray}
where
%Eq.13
\begin{equation}
\lambda(1,u/s,M_\pi^2/s) \equiv \Bigl[1 - \frac{\left(\sqrt{u} + M_\pi\right)^2}{s}\Bigr]
\; \Bigl[1 - \frac{\left(\sqrt{u} - M_\pi\right)^2}{s}\Bigr]\;,
\end{equation}
%Eq.14
\begin{equation}
\lambda(s,u,M_\pi^2) \equiv \Bigl[s - \left(\sqrt{u} + M_\pi\right)^2\Bigr]
\; \Bigl[s - \left(\sqrt{u} - M_\pi\right)^2\Bigr]
\; .
\end{equation}
In the chiral limit the phase space integral $I|_{PS}(s)$ reduces to the simple expression
%Eq.15
\begin {equation}
\lim_{M_\pi^2 \rightarrow 0} I|_{PS}(s) = 3 \,s \;,
\end{equation}
which leads to the well known Pagels \& Zepeda result \cite{PZ}
%Eq.16
\begin{equation}
\frac{1}{\pi}\; Im \;\psi_5(s)|_{\pi\pi\pi} \; = \theta(s) \;\frac{1}{3}\;\frac{M_\pi^4}{f_\pi^2} \frac{1}{2^8\,\pi^4}\; s\;.
\end{equation}
Using Eq.(11) to normalize a model involving two Breit-Wigner forms leads to
%Eq.17
\begin{equation}
\frac{1}{\pi}\; Im \;\psi_5(s)|_{RES} \; = Im \;\psi_5(s)|_{\pi\pi\pi} \frac{[BW_1(s) + \kappa \;BW_2(s)]}{(1+\kappa)} \;,
\end{equation}
where $BW_1(0) = BW_2(0) = 1$, with
%Eq.18
\begin{equation}
BW_i(s) = \frac{M_i^2 (M_i^2 + \Gamma_i^2)}{(s - M_i^2)^2 + M_i^2 \Gamma_i^2}\;\;\;\;(i=1,2)\;,
\end{equation} 
and $\kappa$ is a free parameter controlling the relative weight of the resonances.
Imposing that the area under the spectral function in zero width be equal to the area under the spectral function in finite width, in the limit $\Gamma \rightarrow 0$, will then determine the value of the leptonic decay constant of the first radial excitation, $f_{\pi_1}$, if only the first resonance is used.  In fact, using the relation
%Eq.19
\begin{equation}
\lim_{\Gamma \rightarrow 0} \frac{1}{(s-M_1^2)^2 + M_1^2 \Gamma_1^2} = \frac{1}{M_1 \Gamma_1} \;\pi \; \delta(s-M_1^2)\;,
\end{equation}
gives  for $f_{\pi_1}$ the prediction
%Eq.20
\begin{equation}
f_{\pi_1} = \frac{M_\pi^2}{f_\pi} \frac{1}{16} \left(\frac{M_1}{6\,\pi^3\,\Gamma_1}\right)^{1/2}\;. 
\end{equation}
Notice that the correct chiral behaviour of $f_{\pi_1}$ is obtained, i.e. $f_{\pi_1} = \cal{O}$$ (M_\pi^2)$, as $f_{\pi_1}$ must vanish like $M_\pi^2$ in the chiral limit because $\pi_1$ is not a Goldstone boson (the chiral symmetry remains $SU(2) \times SU(2)$ in the presence of radial excitations of the pion).
Using in Eq.(20) $M_1 \simeq 1.3 \;\mbox{GeV}$, and $\Gamma_1 \simeq 200 - 400\;$  gives
%Eq.21
\begin{equation}
f_{\pi_1} \simeq 2\; \mbox{MeV}\;.
\end{equation}

In order to reduce  the model dependency of the hadronic resonance parametrization it was proposed recently \cite{DNS1}, \cite{DNS2}-\cite{DNS3} to choose an integration kernel in the FESR, e.g. in Eq.(6), such that it vanishes at the peak of each resonance
%Eq.22
\begin{equation}
\Delta_5(s) = 1 \,- \, a_0\, s \,-\, a_1\, s^2 \;,
\end{equation}
so that $\Delta_5(M^2_1) = \Delta_5(M^2_2) =0$, which fixes $a_0 = 0.894\; \mbox{GeV}^2$, and $a_1 = - 0.179 \;\mbox{GeV}^4$. This simple kernel does achieve a substantial reduction of the systematic uncertainty arising in the hadronic sector. In fact, the resonance contribution to the up-, down- and strange-quark  masses is at the level of a couple of percent, well below the error due to the uncertainty in the strong coupling. The optimal choice of integration kernel is most likely application dependent. For instance, pinched kernels in FESR for the vector and axial-vector correlators
\cite{PINCH} improve considerably the agreement between data and the first two Weinberg sum rules. However, a combination of Eq.(22) and pinched kernels does not work as well as Eq.(22) on its own \cite{DNS3}.\\

A generalization of Eq.(22) in the form of a n-th degree Legendre type polynomial has been used in the past to determine meson coupling constants \cite{L1}, heavy quark masses \cite{L2}, and to extract chiral vacuum condensates from data, as well as to determine the counter term of the $\cal{O}$$(p^4)$ CHPT Lagrangian \cite{L3}. We shall also use such a kernel here in order to compare with the simple form, Eq.(22). We introduce the general polynomial
%Eq.23
\begin{equation}
P_{n}(s)= \sum_{m=0}^n\ a_m \; s^{m}\; ,  
\end{equation}
where the coefficients are fixed by (i) imposing a normalization condition at
threshold 
%Eq.24
\begin{equation}
P_{n}\left( s=M_{\pi }^{2}\right) \,=\,1 \;,  \
\end{equation}
and (ii) by requiring that the polynomial $P_{n}(s)$  minimizes the
contribution of the continuum in the range $\left[ s^{(k+1)}_{\mathrm{th}},s_0
\right] $ in a least square sense, i.e.,
%Eq.25 
\begin{equation}
\int_{s^{(k+1)}_{\mathrm{th}}}^{s_0}\,\,s^{m} \, P_{n}(s)\,\,ds=0\,\,\;\;\mathrm{for}%
\,\,m=0,\ldots n-1 \;.  
\end{equation}%
The meaning of the threshold $ s^{(k+1)}_{\mathrm{th}}$ is as follows. The hadronic resonance spectral function can be split into the sum of a term involving the known pionic radial excitations, and a term involving  unknown resonances,  non-resonant background, etc., i.e.
%Eq.26
\begin{equation}
Im \,\psi_5 ^{(k)}(s)\;=\; \sum_{i=0}^{k} Im \,\psi_{5}^{(i)}(s)+\theta (s-s^{(k+1)}_{
\mathrm{th}}) \;Im \,\psi _{5}^{(k+1)}(s)  \;,
\end{equation}
where the upper index $k$ stands for the number of resonances explicitly
included in the sum. The polynomials obtained in this way are closely related to the Legendre polynomials, so that Eq.(25) is exact on account of  orthogonality. Specifically, denoting by $\cal{P}$$_n(x)$ the Legendre polynomials, their relation to the $P_n(s)$ above is
%Eq.27
\begin{equation}
P_{n}(s)\,=\,\frac{\mathcal{P}_{n}\left( x(s)\right) }{\mathcal{P}_{n}\left(
x(M_{\pi }^{2})\right) }  \;,
\end{equation}
where the variable $x(s)$ is
%Eq.28
\begin{equation}
x(s)\,=\,\frac{2s\,-\,(s_0+s^{k+1}_{\mathrm{th}})}{s_0-s^{(k+1)}_{\mathrm{th}%
}}\;,
\end{equation}
defined in the range $x\,\in \,[-1,1]$ for  $s\,\in \,[s^{(k+1)}_{\mathrm{th}%
},s_0]$. The normalization of the $\cal{P}$$_n(x)$ is such that $\cal{P}$$_1(x) = x$, $\cal{P}$$_2(x) = (5 x^3 - 3x)/2$ , etc.

\section{Integration in the complex s-plane}
The FESR  determining $\psi_5(0)$, Eq.(6), can be written as
%Eq.29
\begin{equation}
\psi_5(0)\; \Delta_5(0)  =  2\, f_\pi^2\, M_\pi^2 \, \Delta_5(M_\pi^2) \, + \, \delta_5(s_0)|_{RES}\, +\, \delta_5(s_0)|_{QCD} \;,
\end{equation}
where
%Eq.30
\begin{equation}
\delta_5(s_0)|_{RES} \,= \, \frac{1}{\pi }\int_{9 M_\pi^2}^{s_0} \frac{\Delta_5(s)}{s}\,Im\,\psi _{5}(s)|_{RES} \;ds \;,
\end{equation}

and
%Eq.31
\begin{equation}
\delta_5(s_0)|_{QCD} \,= \,\frac{1}{2\pi i}\oint_{C(|s_0|) } \,\frac{\Delta_5(s)}{s} \;\psi _{5}(s)|_{QCD}
\,ds\,,
\end{equation}
and equivalent expressions for the Legendre type polynomial kernels, where $\Delta_5(s)$ becomes $P_n(s)$ as defined in Eqs. (23)-(28). The contour integral in the complex s-plane can be evaluated in the framework of Fixed Order Perturbation Theory (FOPT), or Contour Improved Perturbation Theory (CIPT)\cite{CIPT}. In FOPT the coupling and the quark masses at a scale $s_0$ are considered fixed (constant) so that  only logarithmic terms contribute to the integral. The renormalization group (RG) summation of leading logs is only carried out after the contour integration by setting $\mu^2 = - s_0$. In this case the integration can be done analytically. In fact, defining
%Eq.32
\begin{equation}
J(q,k)=\frac{1}{2\pi i}\oint_{C|s_0|}s^{q}\,\left( \ln 
\frac{-s}{\mu ^{2}}\right) ^{k}ds \;,
\end{equation}%
where $k=0,1,...5$ and $q=0,1,..,n$, one can derive the master formula 
%Eq.33
\begin{equation}
J(q,k)=s_0^{q+1} \, \sum_{p=0}^k \sum_{l=0}^{k-p} \,  \frac{k!}{p! \, \, l!}  \,
 \frac{1-(-1)^p}{2} (-1)^{\frac{p+1}{2}} \, 
\frac{(-1)^{k+l}}{(q+1)^{k-p-l+1}} \,\pi^{p-1} 
\left( \ln \frac{s_0}{\mu^2} \right)^l \;.
\end{equation}
Explicit expressions for $\delta_5(s_0)|_{QCD}$ may be found in \cite{DNS1}. In the case of CIPT the strong coupling and the quark masses are running and the RG is implemented before integrating. The FESR  then must involve the second derivative of the pseudoscalar correlator which is introduced through the identity
%Eq.34
\begin{equation}
\oint_{C(|s_0|)} ds \, g(s) \, \psi_5(s) \equiv \oint_{C(|s_0|)} ds \, [F(s) - F(s_0)] \;\psi_5''(s) \;,
\end{equation}
where
%Eq.35
\begin{equation}
F(s) = \int ^s ds' \left[ \int^{s'} ds'' g(s'') - \int^{s_0} ds'' g(s'')\right] \;,
\end{equation}
and $g(s)$ is an arbitrary function. This identity is easily proved by integrating by parts the right hand side to obtain the left hand side. 
We choose  $g(s) = \Delta_5(s)/s$, with $\Delta_5(s)$ given in Eq.(22). The function $F(s)$ then becomes
%Eq.36
\begin{equation}
F(s) = s\, ln(s/s_0) - \left(1 \,+\, a_0\, s_0\,+\,\frac{1}{2}\,a_1\,s_0^2\right)\,s + \frac{a_0}{2}\,s^2 \,+\,\frac{a_1}{6} \,s^3 \;.
\end{equation}
The running coupling and the running quark masses can be computed at each step in the integration contour by solving numerically the corresponding RG equations (RGE)\cite{SANTA}-\cite{VERMA}. The running coupling satisfies the equation
%Eq.37
\begin{equation}
s \; \frac{d \, a_s(-s)}{d s} = \beta (a_s) = -  \sum_{N=0} \beta_N \; a_s(-s)^{N+2} \;,
\end{equation}
where $a_s \equiv \alpha_s/\pi$, and for three quark flavours $\beta_0 = 9/4$, $\beta_1 = 4$, $\beta_2 = 3863/384$, $\beta_3 = (421797/54 + 3560 \zeta(3))/256$. In the complex s-plane $s = s_0\, e^{ix}$ with the angle $x$ defined in the interval $x \in (- \pi, \pi)$. The RGE then becomes
%Eq.38
\begin{equation}
\frac{d \, a_s(x)}{d x} = - i \sum_{N=0} \beta_N \; a_s(x)^{N+2} \;.
\end{equation}
%FIG1
\begin{figure}
[ht]
\begin{center}
\includegraphics[height=3.5in, width=5.0in]
{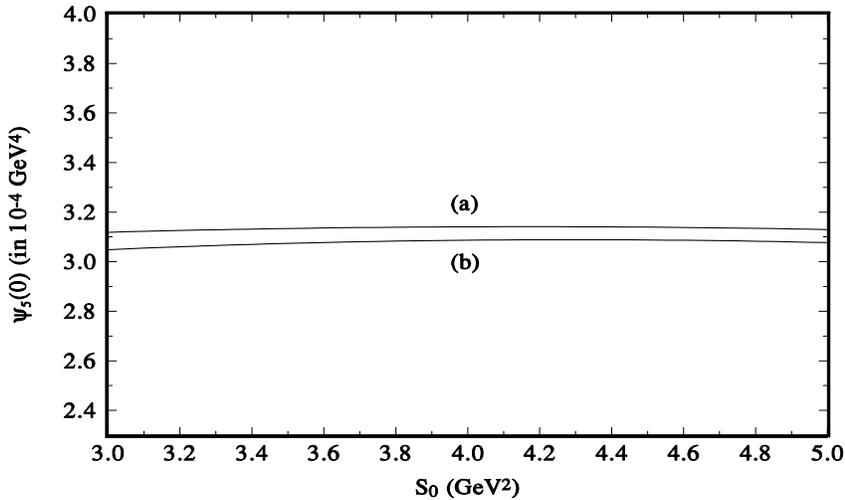}
\caption{Results for $\psi_5(0)$ in units of $10^{-4}\, \mbox{GeV}^4$ as a function of $s_0$ in FOPT and neglecting the hadronic resonance contribution. Curve (a) corresponds to $\alpha_s(M_\tau^2) =0.335$ ($\Lambda\,=\, 365\; \mbox{MeV}$), and curve (b) to $\alpha_s(M_\tau^2) =0.353$
($\Lambda\,=\, 397\; \mbox{MeV}$).}
\end{center}
\end{figure}
This RGE can be solved numerically using e.g. a modified Euler method, providing as input $ a_s (x=0) = a_s (- s_0)$. Next, the RGE for the quark mass is given by
%Eq.39
\begin{equation}
\frac{s}{m} \; \frac{d \, m(-s)}{d s} = \gamma (a_s) = - \sum_{M=0} \gamma_M \; a_s^{M+1} \;,
\end{equation}
where for three quark flavours $\gamma_0 = 1$, $\gamma_1 = 182/48$, $\gamma_2 = [8885/9 - 160 \,\zeta(3)]/64$, $\gamma_3 = [2977517/162 - 148720 \,\zeta(3)/27 + 2160 \,\zeta(4) - 8000\, \zeta(5)/3]/256$. With the aid of Eqs. (37)-(38) the above equation can be converted into a differential equation for $m(x)$ and integrated, with the result
%Eq.40
\begin{equation}
m(x) = m(0) \;exp \Big\{ - i \int_0^x dx' \sum_{M=0} \gamma_M \, [a_s(x')]^{M+1}\Big\}\;,
\end{equation}
where the integration constant $m(0)$ is identified as the overall multiplicative quark mass in the expression for the pseudoscalar correlator, i.e. $m \equiv [m_u(s_0) + m_d(s_0)]$.
%FIG2
\begin{figure}
[ht]
\begin{center}
\includegraphics[height=3.5in, width=5.0in]
{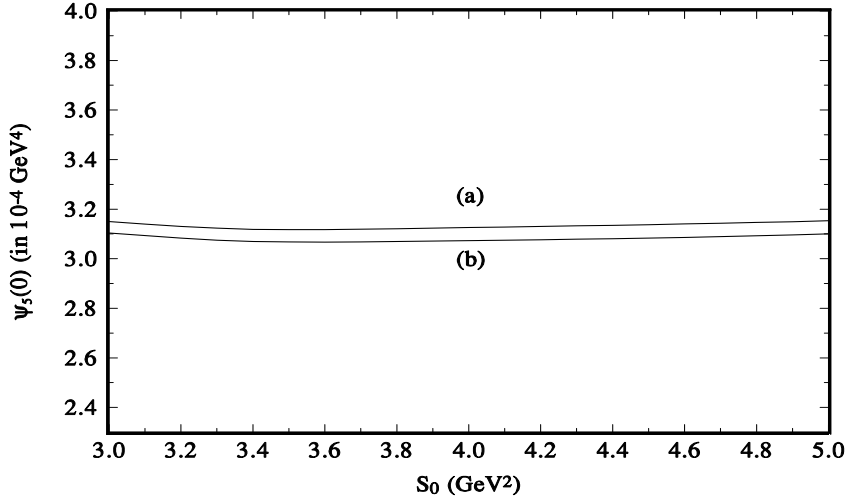}
\caption{Results for $\psi_5(0)$ in units of $10^{-4}\, \mbox{GeV}^4$ as a function of $s_0$ in CIPT and using a two-resonance parametrization of the hadronic spectral function normalized at threshold according to CHPT. Curve (a) corresponds to $\alpha_s(M_\tau^2) =0.335$ ($\Lambda\,=\, 365\; \mbox{MeV}$), and curve (b) to $\alpha_s(M_\tau^2) =0.353$
($\Lambda\,=\, 397\; \mbox{MeV}$).}
\end{center}
\end{figure}
\section{Results}
We begin by discussing the results of the first method, i.e. using the integration kernel Eq.(22). This choice reduces considerably the resonance contribution to $\psi_5(0)$, which becomes smaller than the uncertainty due to the error in the strong coupling, both in FOPT as well as in CIPT. In fact, the difference  between including and not including  the resonance sector is at the level of less than $1\%$ in the result for $\psi_5(0)$. We use the recent high precision determination \cite{ALPHA} of the strong coupling $\alpha_s(M_\tau^2) = 0.344 \,\pm\, 0.009$, which corresponds to a QCD scale in the $\overline{MS}$ scheme of $\Lambda = 365 \,-\, 397\, \mbox{MeV}$. This value is in perfect agreement with the result of a more recent analysis using all present experimental and theoretical knowledge \cite{PICH}, i.e  $\alpha_s(M_\tau^2) = 0.342 \,\pm\, 0.012$; the tiny difference between these two values of the coupling have no impact in our final results.

For the quark masses we use the most accurate determination using the kernel Eq.(22) \cite{DNS3}, essentially free of systematic uncertainties, $\overline{m_u} (2\,\mbox{GeV})\,=\, 2.9\,\pm\, 0.2\, \mbox{MeV}$, and $\overline{m_d} (2 \,\mbox{GeV})\,=\, 5.3\,\pm\, 0.4\, \mbox{MeV}$. The uncertainties are mostly due to  the strong coupling. These running masses correspond to the following invariant masses (the integration constants in the RGE)
$\hat{m_u}\,=\, 3.8\,-\,3.9 \,\mbox{MeV}$ and $\hat{m_d}\,=\, 6.4\,-\,7.2\, \mbox{MeV}$. 
The results for $\psi_5(0)$ in FOPT, and neglecting completely the hadronic resonance contribution is shown in Fig.1 for $\alpha_s(M_\tau^2) =0.335$ ($\Lambda\,=\, 365\; \mbox{MeV}$) (curve(a)), and $\alpha_s(M_\tau^2) =0.353$
($\Lambda\,=\, 397\; \mbox{MeV}$) (curve(b)). These results are fairly stable in the wide region $s_0 = 3.0\,-\, 5.0\, \mbox{GeV}^2$, giving
$\psi_5(0)\,=\, 3.12\,-\,3.14\; \times\,10^{-4}\,\mbox{GeV}^4$ for $\Lambda\,=\, 365\; \mbox{MeV}$, and $\psi_5(0)\,=\, 3.08\,-\,3.09\; \times\,10^{-4}\, \mbox{GeV}^4$ for $\Lambda\,=\, 397\, \mbox{MeV}$. Taking now the resonance contribution into account, and using a two-resonance parametrization normalized to CHPT at threshold, as described in Section 2, $\psi_5(0)$ increases by less than 0.6\% for each value of $\Lambda$. The principal source of error is then  that from $\alpha_s$, from  the $s_0$ dependence, and obviously from the unknown six-loop order in PQCD. Combining these values of $\psi_5(0)$ leads to the following result for $\delta_\pi$
%Eq.41
\begin{equation}
\delta_\pi|_{FOPT} = (6.5 \,\pm \, 0.9) \%.
\end{equation}
%FIG3
\begin{figure}
[ht]
\begin{center}
\includegraphics[height=3.5in, width=5.0in]
{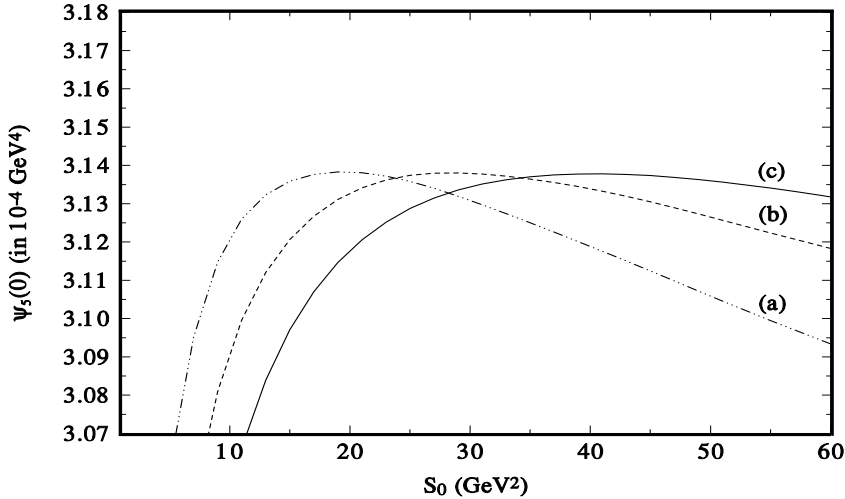}
\caption{Results for $\psi_5(0)$ in units of $10^{-4}\, \mbox{GeV}^4$ as a function of $s_0$ in a fixed $\mu^2 = 4\,\mbox{GeV}^2$ scheme, and for the central value $\alpha_s(M_\tau^2) = 0.344$. Curves (a), (b), (c)  correspond to polynomial degrees n = 4, 5, and 6, respectively.}
\end{center}
\end{figure}
In Fig.2 we show $\psi_5(0)$ obtained in CIPT with the two-resonance parametrization of Section 2 for both values of $\alpha_s$. Once again due to the presence of the integration kernel, the systematic uncertainty due to the resonance contribution is much smaller than  the uncertainty due to the value of $\alpha_s(M_\tau^2)$ and the $s_0$ dependence. The result is $\psi_5(0)\,=\, 3.07\,-\,3.10\; \times\,10^{-4}\, \mbox{GeV}^4$, and $\psi_5(0)\,=\, 3.12\,-\,3.15\; \times\,10^{-4}\, \mbox{GeV}^4$, corresponding, respectively, to  
 $\alpha_s(M_\tau^2) =0.353$
($\Lambda\,=\, 397\; \mbox{MeV}$), and 
$\alpha_s(M_\tau^2) =0.335$
($\Lambda\,=\, 365\; \mbox{MeV}$).
These results translate into
%Eq.42
\begin{equation}
\delta_\pi|_{CIPT} = (7.0 \,\pm \, 0.8) \%,
\end{equation}
in very good agreement with the FOPT result, Eq.(41).\\

Next, we use the Legendre-type integration kernel, Eq.(23), together with the global constraint, Eq.(25). The  pseudoscalar correlator is known to be affected by the rather slow convergence of its PQCD expansion. This is due to the monotonically increasing coefficients multiplying the higher powers of $\alpha_{s}$. As a consequence of this, the convergence of the results will depend crucially on the renormalization prescription being used. We favour a pragmatic \emph{a posteriori} approach. The preferred scheme should be the one that yields the most stable results with respect to two parameters, the degree of the polynomial $P_n(s)$ and the duality radius $s_0$. As far as the renormalization scheme is concerned there are two obvious choices related to the renormalization point in the asymptotic expansion: we can either fix $\mu^{2}=s$ (CIPT) to avoid large logarithmic terms in the perturbative expansion or, since the logarithms entering our analysis are not really large
yet, use a fixed reference scale, say $\mu =2$ $\mbox{GeV}$.  In the case of the Legendre-type kernels we find that a fixed value of $\mu$ provides  better stability. In Fig.3 we show the results for $\psi_5(0)$ corresponding to $n=4,5,6$, and  the central value $\alpha_s (M_\tau^2) = 0.344$. Using values of $\alpha_s$ in the range $\alpha_s(M_\tau^2) = 0.335 - 0.353$ introduces an uncertainty in $\delta_\pi$ of 5\%, and changing the fixed scale $\mu^2$ in the range $\mu^2 = 2 - 50 \,\mbox{GeV}^2$ produces a larger 18\% uncertainty, leading to
%Eq.43
\begin{equation}
\delta_\pi|_\mu = (5.6 \,\pm \, 1.1) \,\% \;,
\end{equation}
compatible with the FOPT value, Eq.(41), as well as the CIPT result, Eq.(42). The dominant uncertainty is that due the $\mu$ dependence, as indicated above. If one were to use somewhat lower values of the strong coupling, e.g. $\alpha_s(M_\tau^2) = 0.322 \pm 0.02$ in FOPT \cite{BAIKOV}, the result would fall within the error band in Eq.(43).  Combining all three results conservatively by using the whole spread of values, rather than averaging and adding errors in quadrature, leads to the final combined result
%Eq.44
\begin{equation}
\delta_\pi = (6.2 \,\pm \,1.6) \,\% \;,
\end{equation}
with an uncertainty about 50 \% smaller than that in Eq.(3) \cite{DNS1}. This value of $\delta_\pi$ is somewhat higher than earlier results  based on PQCD to NLO \cite{RD}: $\delta_\pi = (4.0 \,\pm \,1.0) \%$, and to NNLO \cite{PRADES}:
$\delta_\pi = (3.5 \,\pm \,1.0) \%$, together with model dependent hadronic spectral functions (thus having a far greater impact on the result than in the present determination).
Using the result, Eq.(44), in Eq.(1), together with the recent most accurate quark mass value \cite{DNS3} $\frac{1}{2} (m_u + m_d)|_{2 GeV} = 4.1\,\pm\, 0.2\; \mbox{MeV}$, and assuming
$\langle0|\bar{u} u|0\rangle \simeq \langle0|\bar{d} d|0\rangle \equiv \langle0|\bar{q} q|0\rangle$, we find
%Eq.45
\begin{equation}
\langle0|\bar{q} q|0\rangle|_{2\,\mathrm{GeV}} = (- 267 \pm 5\; \mbox{MeV})^3 \;.
\end{equation}
The result, Eq.(44), implies, using Eq.(2) together with \cite{JAMIN2}  $L^r_8 (\nu_\chi = M_\rho) = (0.88 \pm 0.24) \times 10^{-3}$, the following prediction 
%Eq.46
\begin{equation}
H^r_2(\nu_\chi = M_\rho) = - (5.1 \pm 1.8) \times 10^{-3}\;,
\end{equation}
in reasonable agreement within errors with the estimate \cite{JAMIN2}
$H^r_2(\nu_\chi = M_\rho) = - (3.4 \pm 1.5) \times 10^{-3}\;$, as well as the determination in Eq.(3) \cite{DNS1}. Using, instead, the lattice QCD result \cite{MILC}
$L^r_8 (\nu_\chi = M_\eta) = (0.58 \pm 0.09) \times 10^{-3}$ gives
%Eq.47
\begin{equation}
H^r_2(\nu_\chi = M_\eta) = - (5.7 \pm 2.0) \times 10^{-3}\;.
\end{equation}
In closing, a few remarks should be made concerning the input used for the strong coupling. First and foremost, the issue addressed in this paper is related to physics at scales closer to the $\tau$-lepton mass than to e.g. the $Z$-boson mass. For this reason, we have chosen to use FOPT and CIPT direct determinations of $\alpha_s(M_\tau)$. The three sources used \cite{ALPHA}-\cite{BAIKOV} lead to values of $\delta_\pi$ within the error band in Eq.(44).
Additionally, determinations of $\alpha_s(M_\tau)$ are based on a transparent, model independent procedure which uses as input experimental data on $\tau$-decays. This should be contrasted with determinations of the strong coupling at the scale of the $Z$-boson mass, which involve far more complicated dynamics. 
The latest determination of $\alpha_s(M_Z)$ \cite{ABBATE} gives results which if extrapolated down to the $\tau$-mass would somewhat disagree (at the one standard deviation level) with the direct determinations of $\alpha_s(M_\tau)$. In fact, the authors of \cite{ABBATE} find $\alpha_s(M_Z) = 0.1135 \pm 0.0010$, implying $\alpha_s(M_\tau) = 0.305 \pm 0.008$ to be compared with the range $\alpha_s(M_\tau) = 0.320 - 0.353$ from direct determinations from $\tau$-decay data in FOPT and CIPT. A previous world average at the $Z$-boson scale \cite{BETH} gives the slightly higher value $\alpha_s(M_Z) = 0.1184 \pm 0.0007$, or $\alpha_s(M_\tau) = 0.3186 \pm 0.0058$. The reconciliation between these results and those at the $\tau$-scale is beyond the scope of this work. Assuming that the problem does not lie with the $\tau$-data, our final result, Eq.(44), incorporates the current precise information on $\alpha_s(M_\tau)$ from both FOPT and CIPT.

\section{Conclusions}
In this paper we have performed a direct determination of the light  pseudoscalar correlator at zero momentum, $\psi_5(0)$, using QCD FESR involving polynomial integration kernels tuned to suppress the experimentally unknown hadronic resonance spectral function. This suppression is successfully achieved, as in the determination of the strange quark condensate \cite{DNS1}, and of the light quark masses \cite{DNS2}-\cite{DNS3}. In fact, the uncertainty due to the resonance contribution to the FESR is much smaller than the uncertainty due to $\alpha_s$, and the $s_0$ dependence. Two types of integration kernels have been used in the FESR. First, a simple second degree polynomial vanishing at the peaks of the first two radial excitations of the pion (local constraint). Second, Legendre type polynomials of varying degrees with global constraints. The fact that results from both methods are consistent with each other provides strong support for this procedure.
The integration in the complex s-plane has been performed according to three methods, i.e. FOPT, CIPT, and a fixed renormalization scale scheme. Results from the different integration kernels, and the different integration procedures are consistent with each other and lead to a substantial reduction of the uncertainty in the corrections to the GMOR relation. This uncertainty is mostly due to the value of the strong coupling, the $s_0$ dependence, and to a lesser extent due to the integration procedure and the nature of the integration kernels. The systematic uncertainty from the resonance sector is basically eliminated by these kernels. As a byproduct, the value of the light quark condensate, and the CHPT low energy constant $H^r_2$ have been obtained. 

%**********************************************************************************

\end{document}